\documentclass{elsart1p}                                                                                          
\usepackage{epsfig}                                                                                               
\usepackage{amssymb}                                                                                              
                                                                                                                  
\journal{Physics Letters B}                                                                                       
                                                                                                                  
\begin{document}                                                                                                  
                                                                                                                  
\begin{frontmatter}                                                                                               
                                                                                                                  
\title{Proton spectra from Non--Mesonic Weak Decay of p-shell $\Lambda$--Hypernuclei and evidence for the two--nucleon induced process}                                                                                            
\author[a,b]{M.~Agnello}, \author[c]{A.~Andronenkov}, \author[d]{G.~Beer},                                        
\author[e]{L.~Benussi}, \author[e]{M.~Bertani}, \author[f]{H.C.~Bhang},                                           
\author[g,h]{G.~Bonomi}, \author[i,b]{E.~Botta},                                                 
\author[j,k]{M.~Bregant}, \author[i,b]{T.~Bressani}, \author[i,b]{S.~Bufalino \thanksref{cor1}},                                   
\author[l,b]{L.~Busso}, \author[b]{D.~Calvo}, \author[j,k]{P.~Camerini},                                          
\author[n,c]{B.~Dalena}, \author[i,b]{F.~De Mori}, \author[n,c]{G.~D'Erasmo},                                     
\author[e]{F.L.~Fabbri}, \author[b]{A.~Feliciello}, \author[b]{A.~Filippi},                                       
\author[n,c]{E.M.~Fiore}, \author[h]{A.~Fontana}, \author[u]{H.~Fujioka},                                         
\author[h]{P.~Genova}, \author[e]{P.~Gianotti}, \author[k]{N.~Grion},                                             
\author[e]{O.~Hartmann},  \author[n]{V.~Lenti},                                               
\author[e]{V.~Lucherini}, \author[i,b]{S.~Marcello \thanksref{cor2}}, \author[x]{T.~Maruta},                                       
\author[r]{N.~Mirfakhrai}, \author[s,h]{P.~Montagna}, \author[t,b]{O.~Morra},                                     
\author[u]{T.~Nagae}, \author[p]{D.~Nakajima}, \author[v]{H.~Outa},                                               
\author[e]{E.~Pace}, \author[c]{M.~Palomba}, \author[c]{A.~Pantaleo},                                             
\author[h]{A.~Panzarasa}, \author[c]{V.~Paticchio}, \author[k]{S.~Piano},                                         
\author[e]{F.~Pompili}, \author[j,k]{R.~Rui},                                    
\author[w]{M.~Sekimoto}, \author[n,c]{G.~Simonetti}, \author[w]{A.~Toyoda},                                       
\author[b]{R.~Wheadon}, \author[g,h]{A.~Zenoni}                                                                   
\linebreak
\author{(The FINUDA Collaboration)}                                                                                 
\linebreak         
and                                                                                                                   
\author[zz,b]{G.~Garbarino}                                                                                               
\address[a]{Dipartimento di Fisica, Politecnico di Torino, Corso Duca degli                                       
Abruzzi 24, Torino, Italy}                                                                                        
\address[b]{INFN Sezione di Torino, via P. Giuria 1, Torino, Italy}                                               
\address[c]{INFN Sezione di Bari, via Amendola 173, Bari, Italy}                                                  
\address[d]{University of Victoria, Finnerty Rd., Victoria, Canada}                                               
\address[e]{Laboratori Nazionali di Frascati dell'INFN, via. E. Fermi, 40,                                        
Frascati, Italy}                                                                                                  
\address[f]{Department of Physics, Seoul National University, 151-742 Seoul,                                      
South Korea}                                                                                                      
\address[g]{Dipartimento di Meccanica, Universit\'a  di Brescia, via Valotti 9,                                     
Brescia, Italy}                                                                                                   
\address[h]{INFN Sezione di Pavia, via Bassi 6, Pavia, Italy}                                                     
\address[i]{Dipartimento di Fisica Sperimentale, Universit\'a  di Torino,                                           
Via P. Giuria 1, Torino, Italy}                                                                                   
\address[j]{Dipartimento di Fisica, Universit\'a  di Trieste, via Valerio 2,                                        
Trieste, Italy}                                                                                                   
\address[k]{INFN Sezione di Trieste, via Valerio 2, Trieste, Italy}                                               
\address[l]{Dipartimento di Fisica Generale, Universit\'a  di Torino,                                               
Via P. Giuria 1, Torino, Italy}                                                                                   
\address[n]{Dipartimento di Fisica Universit\'a  di Bari, via Amendola 173,                                         
Bari, Italy}                                          
\address[u]{Department of Physics, Kyoto University, Sakyo-ku, Kyoto Japan}                                       
\address[x]{Department of Physics, Tohoku University, Sendai 980-8578, Japan}                                     
\address[r]{Department of Physics, Shahid Behesty University, 19834 Teheran,                                      
Iran}                                                                                                             
\address[s]{Dipartimento di Fisica Teorica e Nucleare, Universit\'a  di Pavia,                                      
via Bassi 6, Pavia, Italy}                                                                                        
\address[t]{INAF-IFSI, Sezione di Torino, Corso Fiume 4, Torino, Italy}                                           
\address[p]{Department of Physics, University of Tokyo, Bunkyo, Tokyo                                             
113-0033, Japan}                                                                                                  
\address[v]{RIKEN, Wako, Saitama 351-0198, Japan}                                                               
\address[w]{High Energy Accelerator Research Organization (KEK), Tsukuba,Ibaraki 305-0801, Japan}                                    
\address[zz]{Dipartimento di Fisica Teorica, Universit\'a  di Torino, Via P. Giuria 1, Torino, Italy}                                                                                                  
\thanks[cor1]{Corresponding author. Fax: +39~011~6707324; e-mail address:                                         
bufalino@to.infn.it}                                                                                                \thanks[cor2]{at present, on sabbatical leave at Kyoto University, Kyoto, Japan, under
JSPS Program}  
                                                                                                                  
\begin{abstract} 
 
New spectra from the FINUDA experiment of the Non--Mesonic Weak Decay (NMWD) proton kinetic energy for ${\mathrm{^{9}_{\Lambda}Be}}$, ${\mathrm{^{11}_{\Lambda}B}}$,  ${\mathrm{^{12}_{\Lambda}C}}$, ${\mathrm{^{13}_{\Lambda}C}}$, ${\mathrm{^{15}_{\Lambda}N}}$ and  ${\mathrm{^{16}_{\Lambda}O}}$ are presented and discussed along with the published data on ${\mathrm{^{5}_{\Lambda}He}}$ and ${\mathrm{^{7}_{\Lambda}Li}}$.\\ Exploiting the large mass number range and the low energy threshold (15 MeV) for the proton detection of FINUDA, an evaluation of both Final State Interactions (FSI) and the two--nucleon induced NMWD contributions to the decay process has been done. Based on this evaluation, a linear dependence of FSI on the hypernuclear mass number ${\mathrm{A}}$ is found and for the two--nucleon stimulated decay rate the experimental value of $\Gamma_{2}$/$\Gamma_{p}$=0.43$\pm$0.25 is determined for the first time. A value for the  two--nucleon stimulated decay rate to the total decay rate $\Gamma_{2}/\Gamma_{\rm NMWD}$=0.24$\pm$0.10 is also extracted.
\end{abstract}                                                                                                                                      
                                                                                                                  
\end{frontmatter}                                                                                                 
                                                                                                                  
\section{Introduction}                                                                           
The  Non--Mesonic Weak Decay (NMWD) of $\Lambda$--hypernuclei has been studied quite extensively both experimentally and theoretically since the early days of hypernuclear physics \cite{ches,dalitz}. If, on one hand, studies on the structure of hypernuclei provide interesting information on the hyperon--nucleon  strong interaction \cite{tamu}, on the other hand the weak decay of hypernuclei represents the easiest way to get information on the baryon--baryon strangeness--changing weak interactions \cite{alberico,outa}.

In $\Lambda$--hypernuclei the NMWD occurs through processes which involve a weak interaction of the  $\Lambda$ with one or more nucleons. Sticking to the hadronic vertex $\Lambda \rightarrow \pi N$, if the emitted pion is virtual it can be absorbed by the nuclear medium, resulting in a non--mesonic decay of the following types:
\begin{equation}
\Lambda n \rightarrow nn \quad(\Gamma_{n})~,
\label{gamman} 
\end{equation}
\begin{equation}
\Lambda p \rightarrow np \quad(\Gamma_{p})~,
 \label{gammap} 
\end{equation}
\begin{equation}
\Lambda NN \rightarrow nNN \quad(\Gamma_{2})~.
\label{gamma2} 
\end{equation}
The total weak decay rate of a $\Lambda$--hypernucleus is then:
\begin{equation} 
\Gamma_{\rm T} = \Gamma_{\rm MWD} + \Gamma_{\rm NMWD}~,
\label{gamma} 
\end{equation}
where $\Gamma_{\rm MWD}$=$\Gamma_{\pi^{-}} + \Gamma_{\pi^{0}}$ is the total mesonic weak decay rate and $\Gamma_{\rm NMWD}$=$\Gamma_{1} + \Gamma_{2}$, with $\Gamma_{1}$= $\Gamma_{n}+\Gamma_{p}$; the hypernucleus lifetime is $\tau ={\hbar}/ \Gamma_{\rm T}$.  

The channel (\ref{gamma2}), referred to as two--nucleon induced decay and suggested in Ref.~\cite{al91}, can be interpreted by assuming that the pion from the weak vertex is absorbed by a pair of nucleons, correlated by the strong interaction. Note that the non--mesonic processes can also be mediated by the exchange of mesons more massive than the pion. 

Several important experimental advances in NMWD study have been made in recent years, which have established more precise values of the neutron-- and proton--induced decay rates $\Gamma_{n}$ and $\Gamma_{p}$, solving in this way the long--standing puzzle on the $\Gamma_{n}$/$\Gamma_{p}$ ratio \cite{chumi}.

Despite this progress, one should observe that no experimental evidence has been obtained yet of the
two--nucleon stimulated decay, with the exception of two indirect results \cite{bhang,parker}. Various theoretical papers were dedicated to the calculation of the rates and the nucleon spectra for this mechanism \cite{al91,ramos,prc61,alga,bau2,bau,bau3,nucl-th}, predicting a significant contribution to NMWD.
The first calculation of the two--nucleon induced decay rates was performed within a nuclear matter scheme introducing a phenomenological description of the two--particle two--hole polarization propagator \cite{al91}.
This model was improved in Refs.~\cite{ramos,prc61}, where finite nuclei were considered in the local density approximation, and then a microscopic approach was developed \cite{bau2,bau,bau3,nucl-th}.
In Refs.~\cite{al91,ramos,prc61,alga} only the decay channel $\Lambda np \rightarrow nnp$ was considered, while in the microscopic approach of Refs.~\cite{bau2,bau,bau3,nucl-th} all three two--nucleon induced channels, including $\Lambda nn\to nnn$ and $\Lambda pp\to npp$, were taken into account. In particular, a very recent contribution \cite{nucl-th} studied the effects of Pauli exchange terms in the two--nucleon stimulated NMWD, thus giving more reliable evaluations of $\Gamma_{\rm NMWD}$ and $\Gamma_{2}/\Gamma_{\rm NMWD}$.

The FINUDA experiment performed a complete analysis of the proton energy spectra following the NMWD of ${\mathrm{^{5}_{\Lambda}He}}$, ${\mathrm{^{7}_{\Lambda}Li}}$, ${\mathrm{^{9}_{\Lambda}Be}}$, ${\mathrm{^{11}_{\Lambda}B}}$, ${\mathrm{^{12}_{\Lambda}C}}$, ${\mathrm{^{13}_{\Lambda}C}}$, ${\mathrm{^{15}_{\Lambda}N}}$ and  ${\mathrm{^{16}_{\Lambda}O}}$ hypernuclei. Results for ${\mathrm{^{5}_{\Lambda}He}}$, ${\mathrm{^{7}_{\Lambda}Li}}$ and ${\mathrm{^{12}_{\Lambda}C}}$ have already been published~\cite{bufalino}. Exploiting the large systematics, we present here the results of a model--independent analysis of the contributions of Final State Interaction (FSI) and two--nucleon induced decays to the nucleon spectra. This study leads to the experimental indication of the two--nucleon induced decay process.

\section{The experimental and analysis method}
The results reported  in the present Letter have been obtained by analyzing the data collected by the FINUDA experiment from 2003 to 2007 and correspond to an integrated luminosity of $\sim$ 1.2 fb$^{-1}$. The experimental method is briefly described here, while further details are reported in Ref.~\cite{bufalino}. \\FINUDA is a fixed target experiment installed at one of the two interaction points of the DA$\Phi$NE $e^{+}e^{-}$ $\phi$--factory  of Laboratori Nazionali di Frascati (INFN--Italy). A detailed description of the experimental apparatus can be found 
in Refs.~\cite{fnd,fnd2}.\\
The FINUDA spectrometer has a cylindrical geometry and consists of several sensitive layers which are used for particle localization and identification (dE/dx).
From the beam axis outwards, three main parts can be distinguished, as described below.
\linebreak  
\begin{itemize}
\item
{\it The interaction/target region}: it is located at the center of the apparatus. Here the kaons coming from the $\phi$ decay are identified by means of a  barrel of 12 thin scintillator slabs (TOFINO), surrounded by an octagonal array of $Si$ microstrips (ISIM) facing eight target modules.
\item
{\it The external tracking device}: it consists of four layers of position sensitive detectors arranged in a coaxial geometry and immersed in a $He$ atmosphere to minimize the multiple scattering. A decagonal array of $Si$ microstrips (OSIM) faces the target tiles; the next tracking devices are  two octagonal arrays of low mass drift  chambers (LMDC) and a stereo system of straw tubes (ST).
\item
{\it The external time--of--flight detector}: it is a barrel of 72 scintillator slabs (TOFONE) providing signals for the  trigger and time--of--flight measurements.
\end{itemize}
The whole apparatus is placed inside a uniform 1.0 T solenoidal magnetic field.\\
The present analysis was performed on events collected out of three $\mathrm{^{12}C}$ targets ($1.7$ mm thick, mean density $2.265\ g\ cm^{-3}$), two $\mathrm{^{6}Li}$ ones ($90\%$ enriched, $4$ mm thick), two $\mathrm{^{7}Li}$ ones ($4$ mm thick),  two $\mathrm{^{9}Be}$ ones ($2$ mm thick), a $\mathrm{^{13}C}$ one ($99\%$ enriched powder, $10$ mm thick, mean density $0.350\ g\ cm^{-3}$) and a $\mathrm{D_{2}O}$  filled one (mylar walled, $3$ mm thick).\\
The study of the proton spectra following the NMWD  was performed by selecting all the events with two particles emitted in coincidence: a $\pi^{-}$ carrying the information of the hypernucleus formation (whether in its ground state or in a low lying excited one) and a proton coming from the same $K^{-}$ interaction vertex which gives the signature of the  NMWD.\\
Negative pions and protons were identified by means of both the specific energy loss in OSIM and in the LMDC's and, if present, by the mass identification from the time--of--flight system (TOFINO--TOFONE).\\ The $\pi^{-}$'s were selected by requiring tracks identified by four hits ({\it long tracks}), one in each of the FINUDA tracking detectors, whereas for proton identification we required not only {\it long tracks} but also tracks which do not reach the ST system (hereby indicated, for brevity, as {\it short tracks}). The detection threshold on the energy of protons was 15 MeV, a lower limit never reached by previous experiments.\\
By using this multiple particle identification and the described selection criteria, a $\pi^{-}$  momentum resolution of $\Delta p /p\sim 1\%$ FWHM in the momentum region 260--280 MeV/c was achieved.
The measured proton identification efficiency was $90\%$ and the proton energy resolution was $\Delta E / E \sim 2\%$ FWHM at 80 MeV.\\
The data analysis starts with the selection of the $\pi^{-}$  momentum region corresponding to the $\Lambda$--hypernucleus formation in its ground state or in a low lying excited one decaying to the ground state by electromagnetic emission. The details of the intervals determination are exhaustively described in Ref.~\cite{botta}. The momentum and binding energy intervals selected to identify the formation of the different hypernuclei under study are reported in Table~\ref{tab1}. In order to increase the statistics for ${\mathrm{^{12}_{\Lambda}C}}$ the $\pi^{-}$  momentum selection was extended to all the $\Lambda$ bound region instead of including only the ground state~\cite{bufalino}.

\begin{table}[h] 
\begin{center} 
\begin{tabular}{|c|c|c|c|c|} 
\hline 
target & hypernucleus & $p_{\pi^{-}}$ & B.E. & 
references \\ 
       &              &  (MeV/c)  & (MeV) & \\  \hline 
${\mathrm{^{6}Li}}$ & ${\mathrm{^{5}_{\Lambda}He}}$ & 272 -- 278 &  0.63--5.99 & \cite{zim,kam}\\  \hline 
${\mathrm{^{7}Li}}$ & ${\mathrm{^{5}_{\Lambda}He}}$ & 267 -- 273 & -3.02 -- 1.85 & \\  \hline 
${\mathrm{^{7}Li}}$ & ${\mathrm{^{7}_{\Lambda}Li}}$ & 273 -- 279 & 1.85 -- 7.45 & 
\cite{juric,hashim,hyp06} \\ 
\hline
${\mathrm{^{9}Be}}$ & ${\mathrm{^{9}_{\Lambda}Be}}$ & 280 -- 286 & 1.50 -- 7.00 & 
\cite{juric,hashim,noumi2} \\ 
\hline 
${\mathrm{^{12}C}}$ & ${\mathrm{^{11}_{\Lambda}B}}$ & 258 -- 264 & -2.00 -- 2.75 & 
\cite{fnd,hotchi} \\ 
\hline 
${\mathrm{^{12}C}}$ & ${\mathrm{^{12}_{\Lambda}C}}$ & 264 -- 273 & 2.75 -- 14.00 & 
\cite{fnd,hotchi} \\ 
\hline 
${\mathrm{^{13}C}}$ & ${\mathrm{^{13}_{\Lambda}C}}$ & 274 -- 292 & -2.37 -- 14.30 & 
\cite{hashim} \\  \hline 
${\mathrm{^{16}O}}$ & ${\mathrm{^{15}_{\Lambda}N}}$ & 265 -- 270.5 & $\ $ 0.0 -- 
4.90$\ $ & \cite{hashim} \\  \hline 
${\mathrm{^{16}O}}$ & ${\mathrm{^{16}_{\Lambda}O}}$ & 270.5 -- 282 & 4.90 -- 15.40 & 
\cite{hashim} \\  \hline 
\end{tabular} 
\caption{Summary of the momentum and binding energy intervals selected to 
identify the formation of various hypernuclear systems. First column: target 
nucleus; second column: formed hypernucleus; third column: momentum interval; 
fourth column: binding energy 
interval; fifth column: references of previous missing mass spectroscopy 
experiments.} 
\label{tab1} 
\end{center} 
\end{table}

The proton spectra were obtained by selecting all the protons detected in coincidence with a $\pi^{-}$ in the momentum region reported in the third column of Table~\ref{tab1}.\\In the raw $\pi^{-}$ and proton spectra background is mainly due to the $K^{-} ({\mathit np}) \rightarrow \Sigma^{-} {\mathit p}$ absorption on two correlated nucleons in the absorbing nucleus, followed by the in--flight $\Sigma^{-}\to \pi^- n$ decay. Out of the several processes in $K^{-}$ absorption at rest that lead to a continuous spectrum of $\pi^{-}$, this is the only one affecting the region of the bound states of hypernuclei. This background was simulated, reconstructed  with the same reconstruction program used for the real data and subtracted from the raw proton spectra. \\ Another background source for  $\pi^{-}$ and proton coincidence spectra is the $\Lambda_{qf}$ production and decay. However, the protons coming the in--flight $\Lambda_{qf}$ decay do not contribute to the proton spectra coming from the NMWD because they are below the FINUDA proton detection threshold.\\
A crucial point of the analysis is then the acceptance correction of proton spectra; the acceptance, $\epsilon$, was determined by taking into account the apparatus geometry, the efficiency of the FINUDA pattern recognition algorythm, the trigger request and the quality cuts applied to real data.\\
The acceptance function,
$R = 1/\epsilon$, for protons was evaluated target by target in different apparatus sector and for the energy it has a negative quadratic exponential behaviour in the 0-50 MeV range and flattens above 50 MeV.   
A complete account of the analysis procedure for $\mathrm{^5_{\Lambda}He}$, $\mathrm{^7_{\Lambda}Li}$ and $\mathrm{^{12}_{\Lambda}C}$ is given in Ref.~\cite{bufalino}.
\vspace{-0.3cm}
\section{Results and Discussion}
\begin{figure}[htbp] 
\begin{center} 
\includegraphics[angle=90,width=110mm]{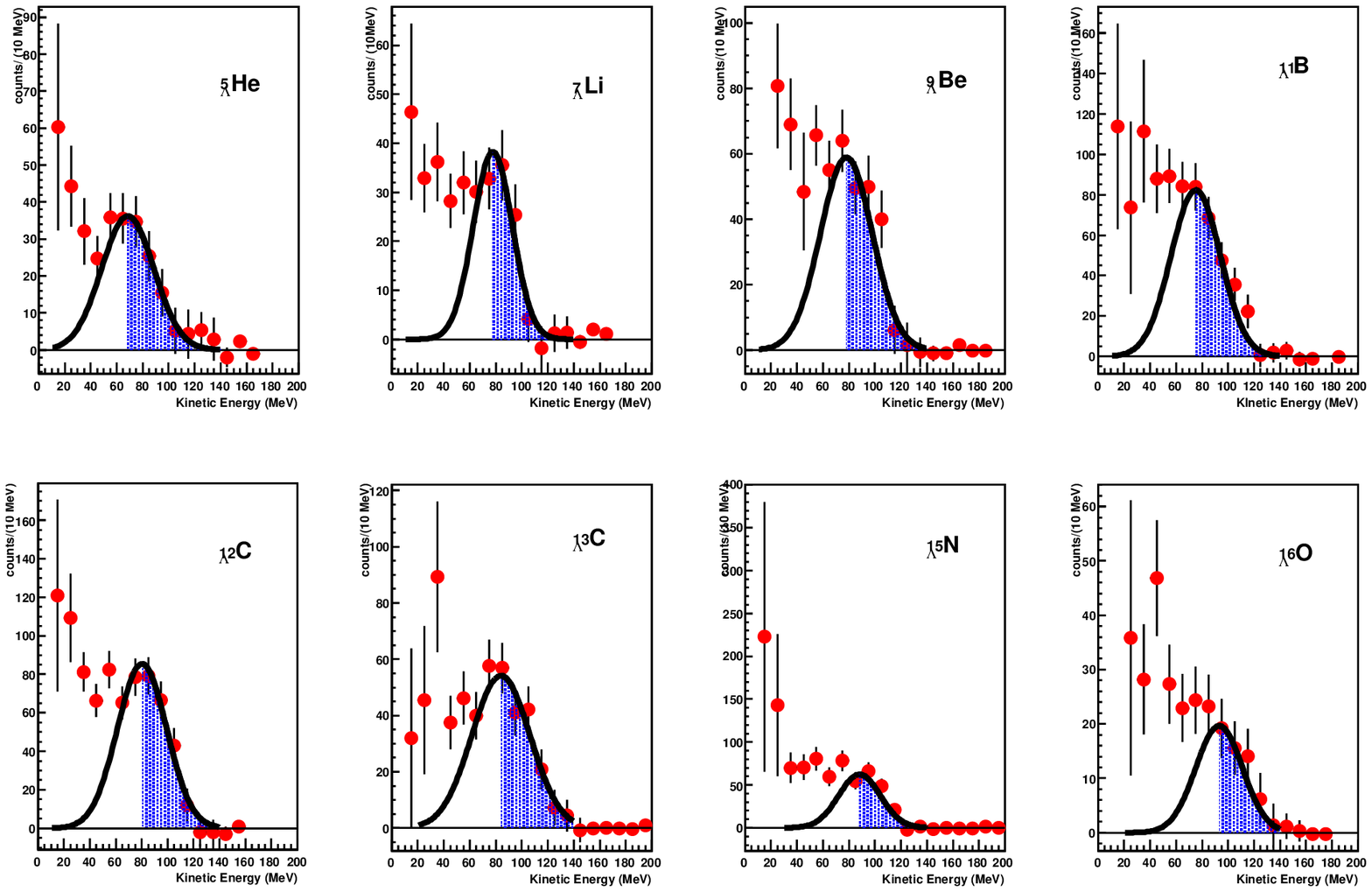} 
\caption{Proton kinetic energy spectra from the NMWD of (from left to right): ${\mathrm{^{5}_{\Lambda}He}}$, ${\mathrm{^{7}_{\Lambda}Li}}$, ${\mathrm{^{9}_{\Lambda}Be}}$, ${\mathrm{^{11}_{\Lambda}B}}$, ${\mathrm{^{12}_{\Lambda}C}}$, ${\mathrm{^{13}_{\Lambda}C}}$, ${\mathrm{^{15}_{\Lambda}N}}$ and  ${\mathrm{^{16}_{\Lambda}O}}$. 
The blue filled area is the spectrum area in which the two--nucleon induced NMWD is negligible.} 
\label{fig1} 
\end{center} 
\end{figure} 
In Fig.~\ref{fig1} the kinetic energy spectra of protons coming from the NMWD of  ${\mathrm{^{5}_{\Lambda}He}}$, ${\mathrm{^{7}_{\Lambda}Li}}$, ${\mathrm{^{9}_{\Lambda}Be}}$, ${\mathrm{^{11}_{\Lambda}B}}$, ${\mathrm{^{12}_{\Lambda}C}}$, ${\mathrm{^{13}_{\Lambda}C}}$, ${\mathrm{^{15}_{\Lambda}N}}$ and  ${\mathrm{^{16}_{\Lambda}O}}$ are shown. All the spectra are background subtracted and acceptance corrected; the errors in the spectra are statistical only and include both the contributions from background subtraction and acceptance correction. We estimated a systematic error of less than $5\%$.\\
The analysis of data collected on several medium mass hypernuclei is a suitable way to extract more information on the NMWD process and is the relevant point of the present Letter.\\
Even though affected by considerable errors, in particular in the low energy region, the proton kinetic energy spectra show a clear trend as a function of the hypernuclear mass number $\mathrm{A}$ (from 5 to 16): a peak around 80 MeV (which corresponds to about half of the $Q$--value for the free $\Lambda p \rightarrow n p$ reaction) is broadened by the Fermi motion of nucleons and more and more blurred as $A$ increases. The peak is smeared, on its low energy side, by a rise that can be ascribed to FSI and two--nucleon induced weak decays.\\
Each proton spectrum from 80 MeV onwards was fitted to a Gaussian function. The mean values of the Gaussian are reported in Table~\ref{tab2} for each hypernucleus. The FWHM of the Gaussian fits is different for each spectrum and this variation is not related to our energy resolution for protons but it is due to the nucleon Fermi motion.\\ 
The Gaussian fits of the proton spectra, shown by the black solid line in Fig.~\ref{fig1}, are centered at energy values which increase with the hypernuclear mass number. 
To predict the mean value of the peak in the proton spectra originating from one--nucleon induced decays is a complicated issue.
 As a matter of fact, in the process:
\begin{equation}
^{A}_{\Lambda}Z \rightarrow n + p + ^{A-2}(Z-1)~,
\label{2N} 
\end{equation}
one has to specify the binding energy of the residual nuclear state $^{A-2}(Z-1)$. A few different cases can occur: a single residual nucleus, in the ground state or in an excited state, or a more complicated system composed of two or more nuclear fragments (possibly in excited states) can be produced by the decay. To our knowledge there are no detailed calculations on this subject. We have verified that, in the two cases of ${\mathrm{^{5}_{\Lambda}He}}$ and  ${\mathrm{^{7}_{\Lambda}Li}}$, the mean values obtained by the fit are close to those obtained by assuming that the residual nucleus $^{A-2}(Z-1)$ is emitted in the ground state ($^{3}$H and  $^{5}$He respectively). The calculated values in such cases are 70 MeV and 79 MeV, respectively.
With increasing ${\mathrm{A}}$, the fitted peak centers become increasingly smaller than the values calculated by means of a three body phase space calculation (more probable value). This circumstance seems to indicate that decays, in which the final nuclear state consists of two or more fragments, are more probable for larger $\mathrm{A}$.

\begin{table}[h] 
\begin{center}
\begin{tabular}{|c|c|} 
\hline 
Hypernucleus &mean value\\\hline 
$\mathrm{^{5}_{\Lambda}He}$& 68.5 $\pm$ 4.1\\\hline 
$\mathrm{^{7}_{\Lambda}Li}$  &  76.7$\pm$5.2\\\hline 
$\mathrm{^{9}_{\Lambda}Be}$ &78.2$\pm$6.2  \\\hline 
$\mathrm{^{11}_{\Lambda}B}$& 75.1$\pm$5.0\\\hline 
${\mathrm{^{12}_{\Lambda}C}}$&  80.2$\pm$2.1\\\hline 
${\mathrm{^{13}_{\Lambda}C}}$&  83.9$\pm$12.8\\\hline 
${\mathrm{^{15}_{\Lambda}N}}$ &88.1$\pm$6.2\\\hline 
${\mathrm{^{16}_{\Lambda}O}}$& 93.1$\pm$6.2\\\hline  

\end{tabular}
\caption{Mean values of the Gaussian fit performed on the proton energy spectra following the NMWD of ${\mathrm{^{5}_{\Lambda}He}}$, ${\mathrm{^{7}_{\Lambda}Li}}$, ${\mathrm{^{9}_{\Lambda}Be}}$, ${\mathrm{^{11}_{\Lambda}B}}$, ${\mathrm{^{12}_{\Lambda}C}}$, ${\mathrm{^{13}_{\Lambda}C}}$, ${\mathrm{^{15}_{\Lambda}N}}$ and  ${\mathrm{^{16}_{\Lambda}O}}$.} 
\label{tab2} 
\end{center} 
\end{table}

Let us now consider the low energy rise in the proton spectra, which we consider to be due to the combined contributions of FSI and two--nucleon induced NMWD.
If it were due to FSI effects, one should naturally expect that the structure at 80 MeV should still exist, but reduced as number of events since a fraction of them would be degraded in energy and contribute to the low energy part of the spectra. By straightforward arguments we may assume that the number of protons suffering a FSI should increase linearly with $\mathrm{A}$ and then the related amount of events in the tail due to FSI.\\
The additional mechanism that could explain the low energy behaviour is a contribution of the two--nucleon induced NMWD; if the energy released in the weak decay is shared by three nucleons, a low energy rise may show up even for hypernuclei as light as ${\mathrm{^{5}_{\Lambda}He}}$. This contribution to the NMWD should be assumed, as first hypothesis, as independent on (or slowly dependent on) $\mathrm{A}$. Theoretical calculations \cite{alberico} support such a constancy of $\Gamma_2/\Gamma_{\rm NMWD}$ for a wide hypernuclear mass range.\\ A way of disentangling between the two mechanisms is that of studying the behaviour of the low energy tail as a fuction of  $\mathrm{A}$ as we will describe in the following.

To estimate the contribution of  two--nucleon induced NMWD we first assume that the part of proton spectrum beyond the peak mean value at about 80 MeV
is due to protons coming from the $\Lambda p \rightarrow n p$ reaction and that the two--nucleon 
induced channel can be neglected in this region. This assumption is confirmed by theoretical calculations, 
which estimated that beyond 70 MeV the protons from the two--nucleon stimulated decay are about 
$5\%$ of the total number of two--nucleon induced decay protons \cite{alga}. 
On the contrary, the part of spectrum below the peak mean value is due  to protons from both $\Lambda p \rightarrow n p$
and two--nucleon induced decays.

As pointed out before FSI also affects the spectra. Let us then denote with $\mathrm{N_p^{\rm FSI-low}}$ ($\mathrm{N_p^{\rm FSI-high}}$) the 
difference between the number of protons detected in the region below (beyond) the peak of the spectrum and 
the (unknown) number of primary protons which originates from one-- and two--nucleon induced weak decays 
in the same region.
Available calculations \cite{alga} predict that $\mathrm{N_p^{\rm FSI-low}}$ is a positive quantity while $\mathrm{N_p^{\rm FSI-high}}$ is negative:
FSI tends to remove protons from the high energy part of the spectrum
while filling the low energy region.

According to the above assumptions and definitions, for the total number of protons $\mathrm{A_{\rm low}}$ 
in the low energy region (below the center of the peak) and 
$\mathrm{A_{\rm high}}$ in the high energy region (beyond the center of the peak; blue filled areas in Fig.~\ref{fig1}) of the spectra  we thus have:
\begin{eqnarray}
\label{as}
A_{\rm low}&=&N(\Lambda p\to np)/2+N(\Lambda np\to nnp)+N_p^{\rm FSI-low}~, \\
\label{as2}
A_{\rm high}&=&N(\Lambda p\to np)/2+N_p^{\rm FSI-high}~,
\end{eqnarray}
where $\mathrm{N(\Lambda p \to np)}$ is the total number of $\Lambda p \to np$ decays 
(which we assume to be evenly distributed  in the Gaussian peak, i.e., half in $\mathrm{A_{\rm low}}$ and half in $\mathrm{A_{\rm high}}$) and
$\mathrm{N(\Lambda np\to nnp)}$ is the total number of $\Lambda np\to nnp$ decays
(which occur only in $\mathrm{A_{\rm low}}$) contributing to the spectra. 
One can also write:
\begin{equation}
\label{number-gamma}
\frac{N(\Lambda np\to nnp)}{N(\Lambda p\to np)}=\frac{\Gamma_{np}}{\Gamma_{p}}
\simeq \frac{\Gamma_2}{\Gamma_{p}}~.
\end{equation}
Note that in Eq.~(\ref{as}) and in the last equality of Eq~.(\ref{number-gamma})
the two--nucleon induced NMWD has been assumed to be dominated by the
$\Lambda np\to nnp$ channel:
$\Gamma_2=\Gamma(\Lambda np\to nnp)+\Gamma(\Lambda pp\to npp)+\Gamma(\Lambda nn\to nnn)\equiv
\Gamma_{np}+\Gamma_{pp}+\Gamma_{nn}\simeq \Gamma_{np}$ 
(the recent microscopical calculation of Ref.~\cite{nucl-th} obtained 
$\Gamma_{np}:\Gamma_{pp}:\Gamma_{nn}=0.83:0.12:0.04$).

\begin{figure}[htbp] 
\begin{center} 
\includegraphics[width=75mm]{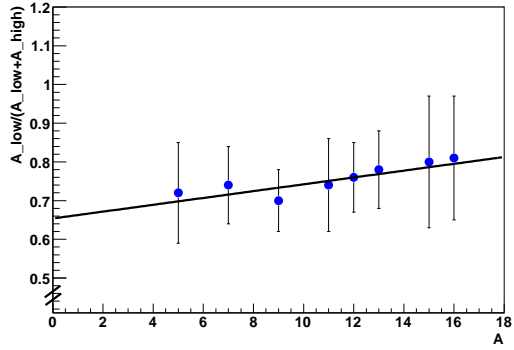} 
\caption{The ratio $A_{\rm low}/(A_{\rm low}+A_{\rm high})$ as a function of the hypernuclear mass number.} 
\label{fig2} 
\end{center} 
\end{figure}

Consider now the ratio:
\begin{equation}
\label{vs-a}
R\equiv \frac{A_{\rm low}}{A_{\rm low}+A_{\rm high}}=\frac{0.5 N(\Lambda p\to np)+N(\Lambda np \to nnp)+N_p^{\rm FSI-{\rm low}}}
{N(\Lambda p\to np)+N(\Lambda np \to nnp)+N_p^{\rm FSI-{\rm low}}+N_p^{\rm FSI-{\rm high}}}~.
\end{equation}
In Fig.~\ref{fig2} the experimental values of this ratio are plotted as a function of $\mathrm{A}$.
Due to FSI effects, both $\mathrm{A_{\rm low}}$ and $\mathrm{A_{\rm high}}$ given by Eqs.~(\ref{as}) and (\ref{as2}) 
are expected to be proportional to $\mathrm{A}$: 
one can thus perform a fit of the data of Fig.~\ref{fig2} with a function
$R(A)=(a'+b'A)/(c'+d'A)$. From this fit it turns out that $d'/c'\leq 10^{-3}$,
thus a linear fit, $R(A)=a+bA$, can be considered.
The result of such a fit ($a= a'/c'=0.654\pm 0.138$, $b= b'/c'=0.009\pm 0.013$ and $\chi^{2}/ndf=0.298/6$) is shown in 
Fig.~\ref{fig2}.\\
Now, we note that we can write $\Gamma_2/\Gamma_p=(\Gamma_2/\Gamma_1)(1+\Gamma_n/\Gamma_p)$. Since theory and experiment support the indipendence of $\Gamma_2/\Gamma_1$ \cite{alberico} and $\Gamma_n/\Gamma_p$ \cite{alberico, bhang} on  $\mathrm{A}$, we may assume that $\Gamma_2/\Gamma_p$ is constant.\\
So one can then rewrite Eq.~(\ref{vs-a}) as:
\begin{equation}
\label{r-a}
R(A)=\frac{\displaystyle 0.5+\frac{\Gamma_2}{\Gamma_p}}
{\displaystyle 1+\frac{\Gamma_2}{\Gamma_p}}+bA~.
\end{equation}

Eq.~(\ref{r-a}) can be solved for $\Gamma_2/\Gamma_p$ for each hypernucleus, getting:
\begin{equation}
 \frac{\Gamma_{2}}{\Gamma_{p}}=\frac{[R(A)-bA]-0.5}{1-[R(A)-bA]}~.
\label{a5} 
\end{equation}
The values obtained for $\Gamma_{2}/\Gamma_{p}$ for our hypernuclei 
are all compatible with each other within errors. The final result can thus be given by
the weighted average:
\begin{equation}
 \frac{\Gamma_{2}}{\Gamma_{p}}=0.43\pm 0.25~.
\label{a6} 
\end{equation}

To determine $\Gamma_{2}$/$\Gamma_{\rm NMWD}$ we need to know the $\Gamma_{n}$/$\Gamma_{p}$ ratio. Indeed:
\begin{equation}
 \frac{\Gamma_{2}}{\Gamma_{\rm NMWD}}=\frac{\Gamma_{2}/\Gamma_{p}}{(\Gamma_{n}/\Gamma_{p})+1 +(\Gamma_{2}/\Gamma_{p})}~.
\label{a8} 
\end{equation}
Theoretical calculations \cite{chumi} predict $\Gamma_{n}$/$\Gamma_{p}$ values between $0.3$ and 
$0.5$ for $^5_\Lambda$He and $\mathrm{^{12}_{\Lambda}C}$ and recent experimental results 
\cite{bhang} give: $\Gamma_{n}/\Gamma_{p}=(0.45\pm 0.11$) for $\mathrm{^5_{\Lambda}He}$ and $\Gamma_{n}/\Gamma_{p}=(0.51 \pm 0.13$) for $\mathrm{^{12}_{\Lambda}C}$. To determine a unique value of  $\Gamma_{n}$/$\Gamma_{p}$ that could be reasonable for all the studied hypernuclei, we take the weighted average of the two cited experimental results, $\Gamma_{n}/\Gamma_{p}=(0.48 \pm 0.08$). 
Using this value together with our determination of $\Gamma_2/\Gamma_p$ we obtain:
\begin{equation}
 \frac{\Gamma_{2}}{\Gamma_{\rm NMWD}}=0.24\pm 0.10~.
\label{a9} 
\end{equation}

One may wonder that the above value~(\ref{a9}) is affected by the experimental threshold of 15 MeV. However it was shown \cite{bau} that proton spectra from two-nucleon induced NMWD, having a three body like phase space distribution, are almost vanishing below 15 MeV. On the contrary the proton spectra from FSI have a monotonic increase going down in energy. Than we expect that the threshold at 15 MeV does not substantially affect the final value of $\Gamma_{2}/\Gamma_{\rm NMWD}$. 
The value derived for the $\Gamma_{2}/\Gamma_{\rm NMWD}$ ratio supports both theoretical predictions \cite{ramos,prc61,alga,bau2,bau,bau3,nucl-th} and the latest experimental results of Ref.~\cite{bhang,parker}, which claim a large contribution of the two--nucleon induced channel to the total NMWD rate. However, the value obtained for  $\Gamma_{2}/\Gamma_{\rm NMWD}$ suggests a smaller contribution than what quoted by Bhang et al. \cite{bhang}, that for $^{12}_{\Lambda}$C reported a contribution of the two--nucleon induced process as large as $40\%$ of the total NMWD width.
We remark that this value was inferred from the fit of a KEK proton spectrum measured with good statistics but poor energy resolution. The KEK spectrum is quite different from the one reported here.
Experimental reasons for such a discrepancy are discussed in Ref.~\cite{bufalino}.

\section {Conclusions}
We reported a systematic study, performed by the FINUDA experiment, of the proton kinetic energy spectra following the NMWD of $\Lambda$--hypernuclei in the $A= 5 \div 16$ mass range. \\
Taking advantage of the low kinetic energy threshold for proton detection (15 MeV) and of the large mass number range of the studied hypernuclei, an evaluation of the nucleon FSI effects and of the two--nucleon induced decay contribution to the decay process was performed. \\ 
A linear dependence of FSI on the hypernuclear mass number $\mathrm{A}$ was found and an experimental value 
of the ratio $\Gamma_{2}/\Gamma_{p}=(0.43 \pm 0.25$) 
was determined for the first time. This result provides a clear experimental indication of the 
relevant r\^ole played by the two--nucleon induced mode in the NMWD of hypernuclei: this channel 
contributes to about 24$\%$ of the $\Gamma_{NMWD}$ for the studied hypernuclei.

\end{document}